\documentclass[conference]{IEEEtran}
\IEEEoverridecommandlockouts
\usepackage[noadjust]{cite}
\usepackage{listings}
\usepackage{tabularx}
\usepackage{tikz}
\usetikzlibrary{shapes.geometric, arrows}
\usetikzlibrary{arrows.meta}
\usepackage{amsmath,amssymb,amsfonts}
\usepackage{algorithmic}
\usepackage{graphicx}
\usepackage{textcomp}
\usepackage{xcolor}
\usepackage{float}
\usepackage{blindtext}
\usepackage{enumitem}
\usepackage{comment}
\usepackage{hyperref}
\usepackage{listings}
\usepackage{verbatim}
\usepackage{colortbl}
\usepackage{multirow} 
\usepackage{array}
\usepackage{makecell}
\usepackage{subcaption}
\usepackage{booktabs}
\usepackage{balance}

\usepackage[listings]{tcolorbox}
\usepackage[skip=0pt, font=small]{caption}
\def\BibTeX{{\rm B\kern-.05em{\sc i\kern-.025em b}\kern-.08em
    T\kern-.1667em\lower.7ex\hbox{E}\kern-.125emX}}
\DeclareCaptionType{mybox}[Box][List of Boxes]

\newboolean{showcomments}
\setboolean{showcomments}{true}
\ifthenelse{\boolean{showcomments}}
{\newcommand{\nb}[2]{
		\fcolorbox{black}{yellow}{\bfseries\sffamily\scriptsize#1}
		{\sf\small$\blacktriangleright$\textit{#2}$\blacktriangleleft$}
	}
	
}
{\newcommand{\nb}[2]{}
	
}

\begin{document}

\title{Detecting Architectural Drift in Safety-Critical Firmware through Runtime Trace Analysis}

\author{
\vspace{-1em}
\hspace*{-1.35em}
\IEEEauthorblockN{
\begin{tabular}{ccc}
\begin{tabular}[t]{c}
Domenico Francesco De Angelis\\
\textit{University of Naples Federico II}\\
\textit{Micron Technology, Inc}\\
Naples, Italy\\
ddeangelis@micron.com\\
domenicofrancesco.deangelis@unina.it
\end{tabular}
&
\begin{tabular}[t]{c}
 Marco De Luca\\
\textit{DIETI}\\
\textit{University of Naples Federico II}\\
Naples, Italy\\
marco.deluca2@unina.it
\end{tabular}
&
\begin{tabular}[t]{c}
 Domenico Amalfitano\\
\textit{DIETI}\\
\textit{University of Naples Federico II}\\
Naples, Italy\\
domenico.amalfitano@unina.it
\end{tabular}
\\[6.3em]
\multicolumn{3}{c}{%
\begin{tabular}[t]{cc}
\begin{tabular}[t]{c}
 Pasquale Cimmino\\
\textit{Micron Technology, Inc}\\
Naples, Italy\\
pcimmino@micron.com
\end{tabular}
\hspace{4em}
&
\begin{tabular}[t]{c}
Anna Rita Fasolino\\
\textit{DIETI}\\
\textit{University of Naples Federico II}\\
Naples, Italy\\
fasolino@unina.it
\end{tabular}
\end{tabular}%
}
\\
\end{tabular}
}
\vspace{-1em}
}

\maketitle

\begin{abstract}
Maintaining consistency between architectural design and runtime-observed behavior is challenging in long-lived safety-critical firmware. This paper presents a runtime-informed methodology for detecting architectural drift in ISO~26262-compliant firmware. The approach collects hardware-assisted execution traces, abstracts them into message exchanges among firmware components, and compares the resulting runtime behavior with design-time sequence diagrams through a deterministic differencing step. The computed delta identifies discrepancies as confirmed, missing, additional, or inverted, while a constrained LLM-based step generates a human-readable report only to support expert review. We evaluate the methodology in an industrial firmware context through agreement-based validation and a practitioner survey. Results over 26 test cases show strong agreement between the generated deltas and expert-curated references, while practitioners perceive the reports as useful for interpreting drift, reducing manual analysis effort, and supporting safety-oriented documentation activities. The findings suggest that combining runtime trace analysis, deterministic architectural differencing, and constrained LLM-based reporting can practically support architectural drift detection in evolving safety-critical firmware.
\end{abstract}

\begin{IEEEkeywords}
Architectural Drift, Architecture Evolution, Software Architecture Recovery, ISO~26262, Embedded Firmware, Execution Trace Analysis
\end{IEEEkeywords}

\section{Introduction}

In automotive firmware development, architecture serves as a design artifact that guides implementation and supports reasoning about system-level concerns, including safety-related decisions. In safety-critical contexts, architectural reasoning primarily relies on components and their interfaces, which constitute the main abstraction used during integration and verification activities~\cite{ISO26262, iso26262-6}. As a result, architectural consistency is often assessed at the level of inter-component interactions rather than internal implementations.

Maintaining architectural documentation aligned with an evolving firmware codebase, however, remains challenging. Continuous evolution driven by feature growth, defect correction, platform variation, and release pressure often causes documented architectures to become partial or misaligned with the deployed implementation~\cite{schmitt2020}. Industrial studies report that while architectural views are crucial for maintainability, communication, and verification, they are difficult to preserve manually over time~\cite{amalfitano2024, zhang2014, altinicsik2024}.

As a result, firmware architecture recovery has become an important industrial activity. Static architecture recovery can reconstruct useful structural views from code, including component decomposition, dependencies, interfaces, and other design-level relations~\cite{amalfitano2024, sartipi2007amalgamated}. However, these views mainly describe what the firmware contains, offering limited evidence on how architectural interactions are actually exercised at runtime, especially when behavior is scenario-dependent, dynamically dispatched, or shaped by event ordering~\cite{sartipi2007amalgamated, huang2006, zellagui2018}.

In ISO~26262-compliant firmware, this gap is critical because architectural confidence depends not only on design intent, but also on evidence that firmware elements interact during execution as prescribed by the design~\cite{ISO26262, iso26262-6, zhang2014}. In this work, we interpret divergences between prescribed and observed message exchanges among firmware components as architectural drift, i.e., runtime-observable deviations between the intended architectural behavior and the behavior actually exercised during execution~\cite{leucker2009brief}. This notion is distinct from architectural erosion, which concerns direct violations of architectural principles~\cite{anthony2024, tekinerdogan2016}. In safety-critical firmware, the distinction is relevant because local and individually acceptable changes may still accumulate into runtime-relevant deviations that affect architectural consistency and complicate verification and integration reasoning~\cite{ali2018, li2022}.

To observe such message-exchange deviations among firmware components, we rely on hardware-assisted instruction tracing~\cite{hendriks2024}. Modern System-on-Chip (SoC) platforms provide dedicated trace units that capture execution non-intrusively, without perturbing real-time behavior~\cite{Stollon2010}. Although implemented differently across architectures, such as ARM ETM~\cite{ARM_ETMv4}, Intel PT~\cite{IntelSDM}, Synopsys RTT~\cite{Synopsys_RTT}, and MIPS PDtrace~\cite{MIPS_PDtrace}, these mechanisms share the principle of recording control-flow changes in a compact stream that can be reconstructed in software.

Building on this context, we present a methodology that acquires firmware traces through a hardware-assisted instruction trace unit and converts them into architectural diagrams of executed behavior. Rather than replacing architecture recovery, the approach complements static architectural views with runtime-informed evidence of actual interactions among firmware components~\cite{zellagui2018}. This dynamic view supports the inspection of interactions and dependencies that may be difficult to infer from code alone and provides practical support for review and compliance-oriented activities in safety-critical settings, where consistency among design, implementation, and verification evidence is central to architectural reasoning~\cite{ISO26262, iso26262-6, amalfitano2024, huang2006}.

The main contributions of this paper are:
(i) a methodology for detecting architectural drift in evolving ISO~26262-compliant firmware by comparing design-time specifications with runtime evidence recovered from execution traces, complemented by a constrained LLM-based reporting step that summarizes the detected drift into a human-readable report;
(ii) an industrial evaluation combining agreement-based validation and a Technology Acceptance Model (TAM)~\cite{davis1989perceived} based practitioner survey.

The rest of the paper is structured as follows:
Section~\ref{sec:survey} introduces the industrial context; 
Section~\ref{sec:proposedProcess} details the methodology and 
Section~\ref{sec:runningExample} illustrates it through running examples; 
Section~\ref{sec:evaluation} reports the evaluation; 
Section \ref{sec:threats} discusses threats to validity;
Section~\ref{sec:lessons} discusses lessons learned; 
Section~\ref{sec:related} covers related work;
and Section~\ref{sec:conclusion} concludes.

\section{Industrial Context and Needs Elicitation} \label{sec:survey}

This section presents the industrial context and the needs elicitation study that motivated the methodology, focusing on challenges in industrial firmware development where architectural documentation must remain aligned with continuously evolving embedded software. Although firmware architecture supports design, integration, and safety-related activities, maintaining accurate descriptions over time remains difficult, especially for legacy and long-lived codebases~\cite{iso26262-8, GAROUSI201814, malladi2016, qa-systems}.

To identify concrete industrial requirements for the proposed methodology, we conducted a qualitative needs elicitation study through focus groups~\cite{krueger_casey} at Micron Technology, the industrial case study of this research.
Six practitioners participated, including firmware developers and firmware testing engineers, with experience in ISO~26262-governed safety systems.

Two authors moderated the sessions using a structured protocol over approximately three hours.
After the sessions, the moderators independently reviewed the focus-group notes, extracted recurring statements, and grouped them into themes related to the use of architectural views, statically recovered models, and runtime traces for reasoning about firmware behavior during execution.
Disagreements in the interpretation or grouping of statements were resolved through discussion. The consolidated themes were then mapped to three concrete industrial needs:

\begin{enumerate}[label={$N_{\arabic*}$}, itemsep=0.2em, topsep=0.2em]
    \item A methodology to identify interaction-level architectural drift from runtime evidence;
    \item A \emph{delta view} that compares a design-time sequence diagram against the corresponding runtime execution trace, making discrepancies between intended and observed component interactions explicit and inspectable, rather than left to manual reconstruction from low-level logs;
    \item A \emph{natural-language report} that describes each detected discrepancy, indicating \emph{what} changed, \emph{where} in the architecture it occurred, and \emph{what type} of change it represents, in order to support practitioners in understanding deviations, reducing the need for direct familiarity with the codebase or the tracing infrastructure.
\end{enumerate}

Overall, these findings highlight a gap between available architectural information and the runtime evidence needed to understand firmware as executed, as also reported in prior industrial studies~\cite{anthony2024, sartipi2007amalgamated, amalfitano2024}. 

\begin{figure}[H]
\centering
\includegraphics[width=1\linewidth]{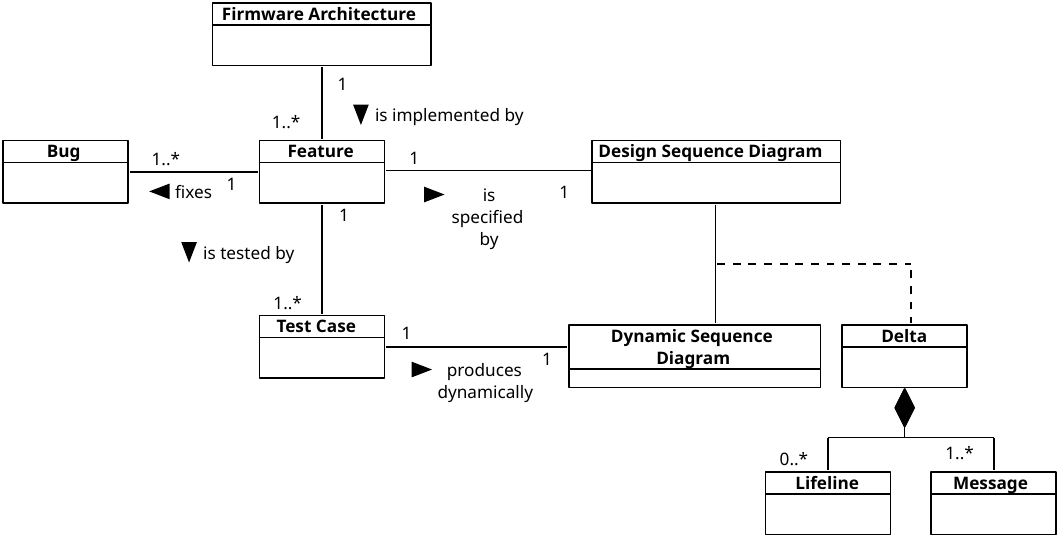}
\caption{Conceptual model of the industrial problem addressed in this work.}
\label{fig:problem}
\end{figure}

Figure~\ref{fig:problem} provides a conceptual view of the industrial problem addressed in this work: firmware architectural artifacts, change drivers, and validation assets evolve in parallel, making it difficult to relate design-time intent to runtime-observed behavior and, consequently, to detect architectural drift.

The \emph{Firmware Architecture} represents the design-time architectural description of the firmware, including its main components, interfaces, and intended interactions. The architecture is implemented by one or more \emph{Features}, which capture planned functional evolutions of the system. During development and validation, \emph{Bugs} may be identified and fixed through features or corrective changes, thereby contributing to the continuous evolution of the firmware.

Each \emph{Feature} is specified by a \emph{Design Sequence Diagram}, which describes the expected interactions among architectural elements for the considered scenario. The same feature is exercised by one or more \emph{Test Cases}. When a test case is executed, it dynamically produces a runtime representation of the observed behavior, modeled as a \emph{Dynamic Sequence Diagram}.

The comparison between the \emph{Design Sequence Diagram} and the corresponding \emph{Dynamic Sequence Diagram} produces a \emph{Delta}. The delta captures discrepancies between intended and observed interactions and represents the basis for architectural drift detection. In the conceptual model, the delta is defined over one or more \emph{Messages} and, when relevant, over \emph{Lifelines}. This reflects the interaction-level nature of the proposed analysis: every delta includes at least one message-level difference, while lifelines are included only when the discrepancy affects the set of participating architectural elements.

\section{Proposed Methodology and Implementation Details}
\label{sec:proposedProcess}

This section presents the proposed methodology and describes how its three phases address the needs identified through the focus-group study. Phase~1 and Phase~2 produce an inspectable delta view between design-time and runtime interactions, thereby addressing $N_2$. Phase~3 generates a human-readable report that describes the detected discrepancies, addressing $N_3$. Together, the three phases support the identification and explanation of message-exchange deviations between design-time specifications and runtime-observed behavior, addressing the overall need expressed in $N_1$.

The methodology proceeds from trace acquisition to drift explanation. Phase~1 extracts, post-processes, and serializes the runtime execution trace as a UML sequence diagram. Phase~2 compares this runtime representation with the design-time sequence diagram and produces a labeled architectural delta. Phase~3 enriches the detected deviations with development-story information from repository mining, serializes the delta, and generates a human-readable report through an LLM-assisted step.

The methodology assumes a strictly sequential execution model, consistent with deterministic scheduling in high-integrity embedded systems~\cite{iso26262-6, kienle2012software}. This is especially relevant for inverted interactions, since ordering differences can be assessed reliably only on a single deterministic execution path~\cite{iso26262-6, Gross2020}. Accordingly, the methodology is scoped to concrete execution scenarios represented as linear sequences of message exchanges.

The following subsections detail each phase, describing inputs, processing steps, and outputs.
\paragraph{Phase 1: Trace Processing}

\begin{figure}[H]
 \centering
 \includegraphics[width=0.85\linewidth]{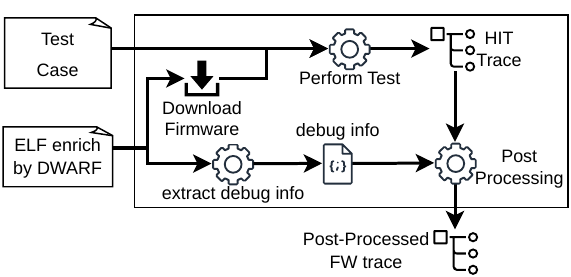}
 \caption{Phase 1 overview.}
 \label{fig:phase1}
\end{figure}

Phase~1 takes as input (i)~the firmware executable in \texttt{ELF} format~\cite{ELF}, enriched by DWARF debugging information~\cite{DWARF5}, and (ii)~a feature-linked test case, i.e., a test case associated with the feature or scenario whose runtime behavior is to be observed, producing a runtime trace in which low-level execution events are resolved against executable-level symbols and related to architectural elements. An overview is shown in Figure~\ref{fig:phase1}.

The \texttt{ELF} binary is analyzed via \texttt{pyelftools}~\cite{pyelftools} to extract \texttt{DWARF} metadata (function boundaries, symbol names, compilation units), building a model that maps runtime addresses to firmware-level entities. The firmware image (ELF files) is then programmed onto the target device (\emph{download firmware}) using a standard firmware flashing procedure, i.e., writing the compiled binary into the device non-volatile memory via a debug or bootloader interface~\cite{openocd,arm_swd}. Once the firmware has been deployed, the selected test case is executed (\emph{Perform Test}) while runtime events are collected as an \emph{HIT Trace}. Collection is deliberately scoped to the execution window of a single test-case execution: this aligns the captured trace with the design-time sequence diagram associated with the exercised feature or scenario, keeps trace volume within HIT bandwidth limits, and ensures that each captured trace corresponds to one concrete execution scenario represented as a linear sequence of observed message exchanges.

The collected \emph{HIT Trace} is post-processed to remove utility functions, redundant low-level repetitions, and execution events irrelevant to the architectural view~\cite{Ducasse2009, sartipi2007amalgamated}. It uses debug information to translate raw addresses into symbol names and modules. The post-processing step does not reconstruct branching or iterative structures from the execution trace. Instead, it converts the observed runtime events into the linear message sequence actually exercised by the selected feature-linked test case. This design choice keeps the runtime representation directly comparable with the design-time sequence associated with the exercised feature or scenario.

The trace is further abstracted by retaining the function invocations that realize runtime message exchanges among architectural entities and, for intra-module calls, only the first invocation level. This abstraction is a deliberate design choice: architectural drift, as adopted in this work, is defined at the level of runtime message exchanges between components. Therefore, both the design-time view (UML sequence diagrams) and the runtime view must be expressed in the same vocabulary of Lifelines and Messages to be commensurable under the Phase~2 delta classification. The resulting \emph{Post-Processed FW trace} retains only the architectural interactions relevant to the observed message-exchange behavior. This trace is subsequently serialized into a PlantUML~\cite{plantuml} sequence-diagram representation for comparison with design-time sequence diagrams.

\begin{figure}[H]
\centering
\includegraphics[width=0.8\linewidth]{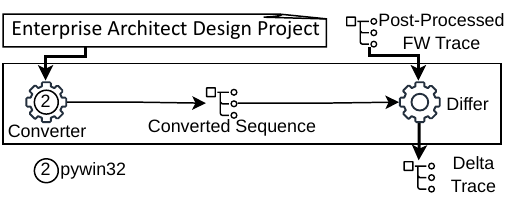}
\caption{Phase 2 overview.}
\label{fig:phase2}
\end{figure}

\paragraph{Phase 2: Delta Computation}
Phase~2 compares design-time and runtime behavioral descriptions in PlantUML format. As input, it takes: (i) the design-time sequence diagrams extracted from UML architectural documentation, and (ii) the processed runtime trace produced in Phase~1. The goal of this phase is to detect drift at the level of inter-component interactions. An overview of this phase is shown in Figure~\ref{fig:phase2}.

In the first step, design-time sequence diagrams are extracted from the \emph{Enterprise Architect} \footnote{Enterprise Architect: \url{https://sparxsystems.com/}} project and converted (\emph{Converter}) into a textual PlantUML format using \texttt{pywin32}~\cite{pywin32}, producing a \emph{Converted Sequence}. This conversion preserves component lifelines and inter-component interactions while removing notation-level differences.

In the second step, the \emph{Converted Sequence} and the \emph{Post-Processed FW trace} produced in Phase~1 are compared by the \emph{Differ}, which computes a diff between the design-time and runtime PlantUML sequences. It is implemented by a Python script that parses both PlantUML files and reduces each message exchange to a triple of source component, target component, and message label.

The classification relies on ordered collections of interaction records. Each record captures the source component, target component, message label, and occurrence within the sequence. 
The \emph{Differ} matches records across the design-time and runtime sequences and classifies them at the architectural message-exchange level.  Interactions present only in the design-time sequence are classified as \emph{missing}, whereas interactions present only in the runtime sequence are classified as \emph{additional}. 
Interactions present in both sequences are classified as \emph{confirmed} when their relative ordering is consistent with the design-time sequence, and as \emph{inverted} when their relative ordering differs. The resulting set of classified message exchanges defines the \emph{Delta Trace}.

The comparison assumes that both inputs describe the same concrete execution scenario as linear sequences of message exchanges. The current implementation does not interpret UML combined fragments, such as \texttt{alt}, \texttt{opt}, and \texttt{loop}, nor does it expand them during delta computation.  Consequently, the method currently applies to design-time sequence diagrams that are already linear. 
Behaviors containing alternative, optional, or repeated interactions would require separate linearized design-time sequences and corresponding feature-linked test cases before being compared by the \emph{Differ}.

\begin{figure}[H]
\centering
\includegraphics[width=0.8\linewidth]{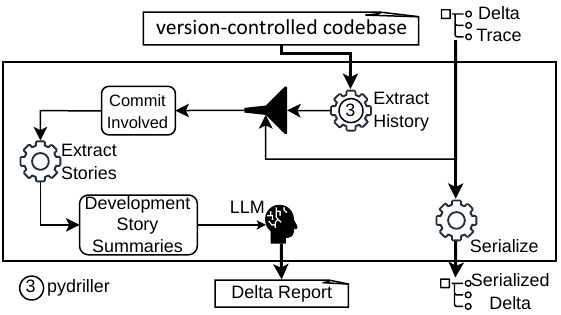}
\caption{Phase 3 overview.}
\label{fig:phase3}
\end{figure}

\paragraph{Phase 3: LLM-Assisted Reporting}
Phase~3 takes as input: (i)~the \emph{Delta Trace} produced in Phase~2, and (ii)~the \emph{version-controlled codebase}. The goal is to produce a readable summary of the detected differences and to serialize the delta. An overview of this phase is shown in
Figure~\ref{fig:phase3}.

As a first step, the \emph{Extract History} process analyzes the repository history to extract the commit history, starting from the first commit corresponding to the date on which the design artifacts were introduced, up to the latest revision in the primary development line. This analysis is performed using \texttt{PyDriller}~\cite{spadini2019pydriller}. The extracted commit history is then filtered by retaining only changes affecting messages appearing in the \emph{Delta Trace}, producing the set of \emph{Commit Involved}. The \emph{Commit Involved} set is processed by the \emph{Extract Stories} step, which correlates the commit history with work-item metadata retrieved from a project management system, producing the \emph{Development Story Summaries}. This enables traceability between architectural deviations and their originating change requests. In compliance with ISO~26262 requirements, development branches and associated work items must be traceable~\cite{ISO26262}.

In parallel, the \emph{Delta Trace} is passed to the \emph{Serialize} step, which produces the \emph{Serialized Delta} in PlantUML format. Before communicating the delta to engineers, an LLM-based post-processing step is applied exclusively at the reporting level. The LLM operates solely on the \emph{Serialized Delta} and the \emph{Development Story Summaries}, producing the \emph{Delta Report}. The LLM does not influence the derivation of architectural relations, or the delta computation; its role is limited to improving interpretability of the computed differences~\cite{lee2025can, chen2025llms}. The generated report serves solely as a review aid to facilitate human inspection of the computed delta, and does not constitute primary safety evidence. All safety-related conclusions remain the responsibility of qualified engineers, in accordance with ISO~26262.

The interaction with the LLM is governed by a structured prompt that constrains the model to operate solely on the serialized delta and development story summaries. The prompt is defined in Box~\ref{box:prompt}.

The outputs of Phase~3 are: (i) an architectural delta in PlantUML (\emph{Serialized Delta}) and (ii) a human-readable delta report (\emph{Delta report}) generated with LLM assistance, providing engineers with structured support for reviewing architectural consistency.

\begin{mybox}
    \caption{Role-constrained prompt for architectural drift report.}
    \centering
    \fbox{\includegraphics[width=0.79\linewidth]{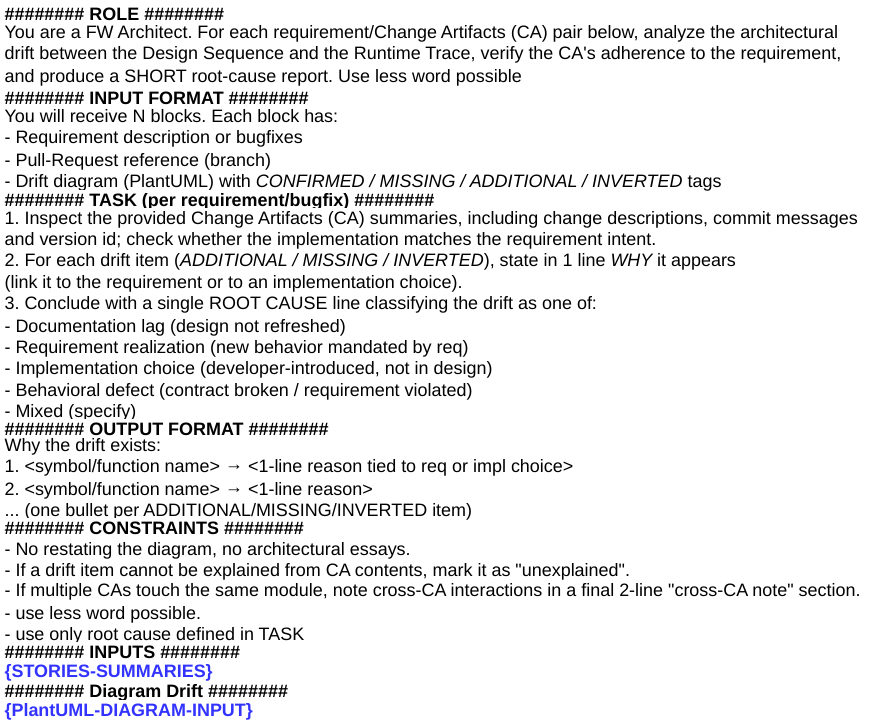}}
    \label{box:prompt}
\end{mybox}

\section{Categories of Architectural Drift}
\label{sec:runningExample}

This section presents the architectural drift categories considered by the proposed methodology when comparing design-time architectural behavior with runtime evidence. We distinguish four cases: \textsc{confirmed}, when the intended interaction is observed at runtime; \textsc{additional}, when runtime evidence contains interactions not specified at design time; \textsc{missing}, when specified interactions are not observed at runtime; and \textsc{inverted}, when the runtime order of interactions differs from the design-time specification.
While \textsc{confirmed} represents the reference case in which no drift is detected, the remaining categories capture different manifestations of architectural drift in component interactions.
The following examples use synthetic component names and interactions for illustrative purposes only.

\subsection{Confirmed interaction}
\begin{figure}[H]
\centering
\includegraphics[width=0.4\linewidth]{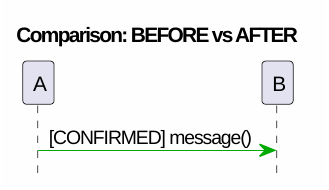}
\caption{Confirmed Interaction: delta view.}
\label{fig:ex1_compare}
\end{figure}

An interaction is classified as CONFIRMED when the same source component, target component, and message label are present in both the design-time and runtime sequences and its relative ordering is consistent with the design-time sequence (Figure~\ref{fig:ex1_compare}), providing evidence that the intended relation is preserved in the executed firmware, in terms of messages and lifelines.

\subsection{Additional interaction}
\begin{figure}[H]
\centering
\includegraphics[width=0.85\linewidth]{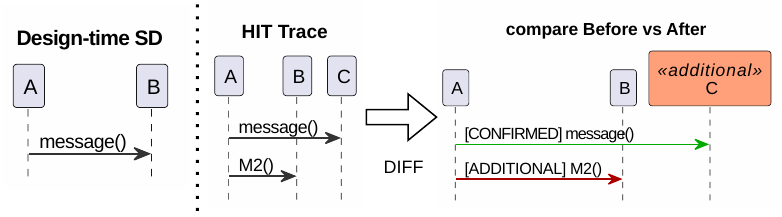}
\caption{Additional interaction: design-time, runtime, delta.}
\label{fig:ex2_compare}
\end{figure}

An \textsc{Additional} interaction arises when the runtime sequence contains a message exchange absent from the design-time specification, as shown in Figure~\ref{fig:ex2_compare}. This deviation may indicate that a component has acquired an unplanned dependency, potentially violating the isolation assumptions required for safety verification under ISO~26262~\cite{iso26262-6}.

\subsection{Missing interaction}
\begin{figure}[H]
\centering
\includegraphics[width=0.85\linewidth]{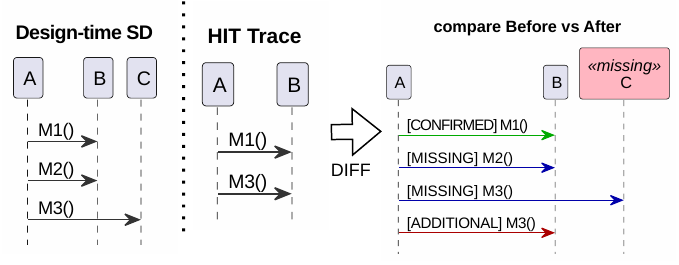}
\caption{Missing interaction: design-time, runtime, delta.}
\label{fig:ex3_compare}
\end{figure}

In contrast to the previous case, a \textsc{Missing} interaction is one that was specified at design time but is not observed in the runtime trace (Figure~\ref{fig:ex3_compare}). The absence may reflect a behavioral regression, 
an untriggered conditional path, or a specification that has become inconsistent with the evolved implementation. Regardless of its root cause, the deviation warrants explicit review: unexercised architectural relations reduce the completeness of verification evidence required by ISO~26262~\cite{iso26262-6}.

\subsection{Inverted interaction}
\begin{figure}[H]
\centering
\includegraphics[width=0.85\linewidth]{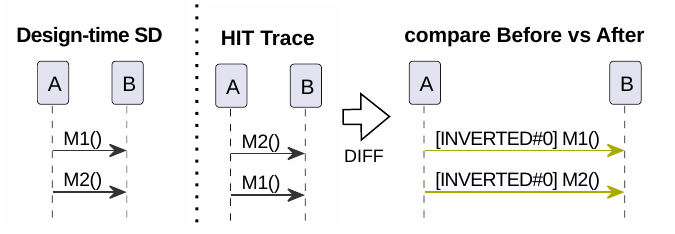}
\caption{Inverted interaction: design-time, runtime, delta.}
\label{fig:ex4_compare}
\end{figure}

Figure~\ref{fig:ex4_compare} shows a scenario in which the runtime interaction order differs from the design-time specification, classified as \textsc{Inverted}.
This emerges when ordering constraints among architectural interactions were not explicitly or correctly captured at design time, possibly revealing implicit dependencies introduced during firmware evolution~\cite{perry1992foundations, li2022, anthony2024}.

\section{Experimental Evaluation}
\label{sec:evaluation}

This section presents the experimental procedure conducted to evaluate the proposed methodology in an industrial firmware setting.

\subsection{Goal and Research Questions}
The goal of the evaluation is to assess the correctness of the detected architectural deltas, the faithfulness of the LLM-generated reports, and the perceived practical value of the methodology in ISO 26262-governed firmware development.

The evaluation is structured around three research questions:

\begin{enumerate}[label={$RQ{\arabic*}.$}, itemsep=0.2em, topsep=0.2em]
    \item \emph{To what extent does the generated
    architectural-drift delta align with a manually curated reference derived from the original sequence diagram and the corresponding trace?} \\
    \textit{Rationale:} Evaluates the correctness of the drift-detection mechanism.

    \item \emph{To what extent is the LLM-generated report faithful in identifying the root cause of architectural drift?} \\
    \textit{Rationale:} Assesses whether the report correctly explains the root cause underlying the detected architectural drift, based on independent expert annotations.
    \item \emph{To what extent is the proposed methodology perceived as useful by practitioners for architectural drift interpretation and ISO~26262 compliance?}\\
    \textit{Rationale:} Examines the perceived practical value for safety-related documentation and evidence generation.
\end{enumerate}

To make explicit how each research question is evaluated, Table~\ref{tab:rq_metrics_subjects} summarizes the metrics and evaluation subjects associated with each RQ.

\begingroup
\begin{table}[t]
\centering
\small
\caption{Mapping between research questions, metrics, and evaluation subjects.}
\label{tab:rq_metrics_subjects}
\begin{tabularx}{\linewidth}{p{0.05\linewidth} p{0.345\linewidth} X}
\toprule
\textbf{RQ} & \textbf{Metrics} & \textbf{Evaluation subjects} \\
\hline
RQ1 & Precision, Recall, F1 & Three senior professionals with expertise in firmware architecture and safety verification \\ \midrule
RQ2 & Fleiss' Kappa & Three senior professionals with expertise in firmware architecture and safety verification \\ \midrule
RQ3 & Likert-scale ratings on PU, PEOU, and compliance support & 23 industry practitioners involved in the anonymous questionnaire \\ 
\bottomrule
\end{tabularx}
\vspace{0.5em}
\small PU = Perceived Usefulness; PEOU = Perceived Ease of Use;
\end{table}
\endgroup

\subsection{Metrics}
\label{sec:metrics}

As summarized in Table~\ref{tab:rq_metrics_subjects}, the evaluation adopts different metrics according to the objective of each research question.

For RQ1, Precision, Recall, and F1-score quantify the alignment between generated deltas and the expert-curated reference~\cite{schutze2008introduction}. The metrics are computed by comparing the elements automatically reconstructed by the methodology with the corresponding ground-truth elements, considering both lifelines and messages. For RQ2, we use Fleiss' Kappa ($\kappa$)~\cite{landis1977measurement} to quantify the agreement among the three expert annotators. 
The annotators independently assessed whether the LLM-generated report faithfully identified the root cause of the detected architectural drift. For RQ3, we rely on the metrics derived from an anonymous questionnaire administered to 23 industry practitioners. 
The questionnaire includes both 5-point Likert-scale items and open-ended questions. The Likert-scale items are used to assess practitioners' perceptions of the generated artifacts across three dimensions: \textit{Perceived Usefulness} (PU), \textit{Perceived Ease of Use} (PEOU), and \textit{Compliance Support} (C). The open-ended questions complement these quantitative ratings by capturing recurring perceived benefits and improvement areas.

\subsection{Objects and Subjects}
The \emph{objects} of the study are internally developed firmware modules representative of industrial embedded systems, evaluated through feature-linked test cases exercising specific architectural scenarios. Due to organizational confidentiality constraints, the evaluation is limited to proprietary firmware modules. The evaluation covered $26$ feature-linked test cases, each involving on average $9$ lifelines and approximately $25$ inter-component messages, reflecting a representative level of industrial architectural complexity. 
Consistent with the scope of the current implementation, the selected feature-linked test cases were associated with linear design-time sequence diagrams, i.e., diagrams not containing UML combined fragments such as \texttt{alt}, \texttt{opt}, and \texttt{loop}. This does not imply that the firmware lacks conditional or iterative logic, but only that such logic was not represented through combined fragments in the evaluated sequence diagrams.

The \emph{subjects} of evaluation relied on two distinct cohorts, as shown in Table \ref{tab:rq_metrics_subjects}.
To investigate RQ1 and RQ2, three senior professionals with backgrounds in firmware architecture and safety verification, not involved in the methodology's development, independently validated the output against a manually curated reference.
Furthermore, for RQ3, a voluntary and anonymous questionnaire was administered to a \textit{purposive sample} of $23$ industry practitioners. The cohort included $14$ embedded firmware developers, $8$ testing and verification engineers, and $1$ software architect. The sample is experienced: $18$ participants reported over two years of professional experience, of whom $10$ exceeded five years, and $16$ reported spending more than $20\%$ of their time on design-related activities, ensuring a relevant and informed evaluation perspective.

\subsection{Experimental Procedure}
\label{sec:procedure}

The experimental procedure was organized according to the three research questions. 
For RQ1, the objective was to assess the capability of the proposed methodology to automatically reconstruct sequence diagrams consistent with the actual implementation. 
For RQ2, the objective was to assess the faithfulness of the LLM-generated reports in explaining the root cause of the detected architectural drift, as evaluated through independent expert annotations. 
For RQ3, the objective was to evaluate the perceived practical value of the generated artifacts from the perspective of industrial practitioners.
\color{black}

\subsubsection{Experimental Procedure for RQ1}
\label{sec:procedure-rq1}

The procedure for RQ1 consisted of the following two steps:

\noindent \textbf{i) Ground-Truth Construction}
Three annotators independently reconstructed the reference sequence diagram for each evaluated feature-linked test case by manually analyzing the source code. The reconstruction focused on the architectural elements represented in the sequence diagram, namely lifelines and exchanged messages. 
Each annotator produced an independent version of the expected sequence diagram, without relying on the sequence diagram automatically generated by the proposed methodology.
After the independent reconstruction, the three resulting diagrams were compared. 
Whenever discrepancies emerged among the annotators regarding specific elements, such as the presence of a lifeline, the presence of a message, or the ordering of a message, the final decision was made through majority voting.  This process produced a reference sequence diagram for each evaluated feature-linked test case, which was used as the ground truth in the subsequent analysis.

\noindent \textbf{ii) Data Analysis}
Three annotators compared each ground-truth sequence diagram with the corresponding sequence diagram automatically reconstructed by the proposed methodology. 
The comparison was performed at the level of individual diagram elements, considering both lifelines and messages.
Each element was assigned to one of the categories defined in Section~\ref{sec:runningExample}, namely \textsc{confirmed}, \textsc{missing}, \textsc{additional}, or \textsc{inverted}. 
The resulting classifications were used to compute quantitative metrics for assessing the reconstruction capability of the proposed methodology, including precision, recall, and F1-score.

\color{black}

\subsubsection{Experimental Procedure for RQ2}
\label{sec:procedure-rq2}

The procedure for RQ2 evaluated the faithfulness of the root-cause explanations provided in the LLM-generated reports. 
In this context, faithfulness refers to the extent to which the report correctly identifies the actual root cause of the detected architectural drift. Three annotators independently analyzed the reports generated by the LLM. For each report, they inspected the explanation of the detected architectural drift and assessed whether the root cause identified by the LLM was correct with respect to the actual cause determined from the technical evidence. Each report was therefore classified according to whether the LLM-generated explanation was faithful or not faithful to the real root cause of the drift. To assess the consistency of the independent judgments, we computed Fleiss' Kappa ($\kappa$) over the annotations provided by the three annotators.

\color{black}

\subsubsection{Experimental Procedure for RQ3}
\label{sec:procedure-rq3}

To assess the practical value of the proposed methodology, we designed a structured survey inspired by the Technology Acceptance Model (TAM)~\cite{davis1989perceived}. 
TAM is widely used to study the adoption of technologies and methods, especially through constructs such as \textit{Perceived Usefulness} (PU) and \textit{Perceived Ease of Use} (PEOU). 
In our study, we also considered \textit{Compliance Support} as an additional construct relevant to the industrial context of architectural documentation and firmware verification.

Participants were presented with the generated artifacts, namely the delta trace and the natural-language delta report. 
They were then asked to evaluate them through the survey items reported in Table~\ref{tab:questionnaire}. 
The questionnaire included both 5-point Likert-scale questions (\emph{1 = strongly disagree (SD)}, \emph{2 = disagree (D)}, \emph{3 = neutral (N)}, \emph{4 = agree (A)}, \emph{5 = strongly agree (SA))} and open-ended questions.
Closed-ended answers were summarized by grouping the responses according to the Likert-scale values. Open-ended answers were analyzed qualitatively by two authors, who identified recurring themes in the participants' comments. A third author subsequently reviewed and validated the resulting thematic interpretation.
\color{black}

\begin{table*}[t]
\centering
\small
\renewcommand{\arraystretch}{0.3}
\caption{Questionnaire items used to answer RQ3 by assessing the perceived practical value of the generated artifacts.}
\label{tab:questionnaire}
\begin{tabularx}{\textwidth}{p{0.2cm} X r r}
\toprule
\textbf{ID} & \textbf{Question} & \textbf{Semantic} & \textbf{Type} \\
\midrule
Q1 &
The report generated by the LLM facilitates understanding of the root cause of architectural drift.
& PU & L \\
Q2 &

The report information was easy to relate to the evaluated feature-linked test case.
& PEOU & L \\
Q3 &
\textit{Did you notice any discrepancies or hallucinations in the report compared to the technical delta?}
& PU & Q \\
Q4 &
The proposed methodology speeds up the identification of design-runtime discrepancies.
& PU & L \\
Q5 &
The methodology provides sufficient documentary evidence to support an ISO~26262 safety audit.
& C & L \\
Q6 &
I find the language and terminology used in the LLM-generated report to be clear and unambiguous.
& PEOU & L \\
Q7 &
Identifying architectural drift through message-exchange deltas is intuitive and easy to follow.
& PEOU & L \\
Q8 &
Overall, the system's interface/output is easy to use.
& PEOU & L \\
Q9 &
If integrated into a CI/CD pipeline, I would use this methodology on a regular basis.
& PEOU & L \\
Q10 &
\textit{What do you consider the most useful aspect of the report generated by the LLM agent?}
& PU & Q \\
Q11 &
\textit{Which feature would you add to better align this methodology with ISO~26262 compliance needs?}
& C & Q \\
\bottomrule
\end{tabularx}
\vspace{0.2em}
\footnotesize L = 5-point Likert scale; Q = Open-Ended Questions; PU = Perceived Usefulness; PEOU = Perceived Ease of Use; C = Compliance Support
\end{table*}

\subsection{Results for RQ1}
\label{sec:rq1}

The results for RQ1 are summarized in terms of precision, recall, and F1-score across the considered categories. 
For confidentiality reasons, we report only aggregate percentage-based metrics and do not disclose absolute counts, such as the number of drift instances identified for each category, since these values could reveal information about the internal quality of the analyzed firmware.

The \textsc{confirmed} class achieved the highest performance, with precision equal to 96.3\,\%, recall equal to 92.9\,\%, and F1-score equal to 94.6\,\%. 
This result indicates that the methodology is particularly effective in correctly reconstructing interactions that are actually present in the ground truth, and therefore in identifying architectural behavior that is preserved with respect to the reference sequence diagram. The \textsc{missing} and \textsc{additional} categories also obtained strong results, with F1-scores of 88.2\,\% and 86.4\,\%, respectively. 
These results show that the methodology can effectively identify both elements that should have been reconstructed but are absent from the generated diagram and elements that are generated by the methodology but are not present in the ground truth. 
Therefore, the approach is able to capture both omissions and unexpected reconstructed interactions. The \textsc{inverted} class obtained lower performance, with precision, recall, and F1-score equal to 60.0\,\%. 
This result suggests that detecting ordering discrepancies between the ground truth and the reconstructed diagram is more challenging than detecting the presence or absence of lifelines and messages. 
This is also consistent with the fact that inverted interactions represent a more specific type of discrepancy, valid only for messages, where each misclassification has a stronger impact on the final metric values.

\begin{tcolorbox}[colback=gray!10,boxrule=0.5pt,title=RQ1 Answer,boxsep=1pt,left=1pt,right=1pt,top=1pt,bottom=1pt]
The proposed methodology shows strong reconstruction capability, achieving high precision, recall, and F1-score for confirmed interactions and strong F1-scores for missing and additional elements. 
The lower performance on inverted messages indicates that ordering-related discrepancies remain the most challenging category to detect.
\end{tcolorbox}
\color{black}

\subsection{Results for RQ2}
\label{sec:rq2}

Following the protocol described in Section~\ref{sec:procedure}, the expert assessment yielded Fleiss'~$\kappa = 0.692$, corresponding to \emph{substantial} agreement according to the interpretation guidelines by Landis and Koch~\cite{landis1977measurement}. 

These results indicate that, in most cases, the LLM-generated reports correctly identified the root cause of the detected architectural drift. 
At the same time, the non-negligible proportion of \emph{not faithful} explanations shows that root-cause identification remains more challenging than describing the presence of drift itself. 
Disagreements among annotators were mainly concentrated in borderline cases, where the LLM provided a partially plausible explanation but did not fully capture the actual technical cause of the drift, or where the available evidence left room for interpretation.

Overall, the substantial agreement among annotators supports the consistency of the expert-based evaluation, while the observed non-faithful cases show that the LLM still struggles to infer the correct root cause in some borderline cases.

\begin{tcolorbox}[colback=gray!10,boxrule=0.5pt,title=RQ2 Answer,boxsep=1pt,left=1pt,right=1pt,top=1pt,bottom=1pt]
The evaluation indicates that LLM-generated reports are generally faithful in identifying the root cause of architectural drift, with substantial agreement among expert annotators. 
However, root-cause explanation remains challenging in borderline cases where the technical evidence is incomplete, indirect, or open to interpretation.
\end{tcolorbox}

\subsection{Results for RQ3}
\label{sec:rq3}

\begin{figure}[H]
    \centering
    \includegraphics[width=1\linewidth]{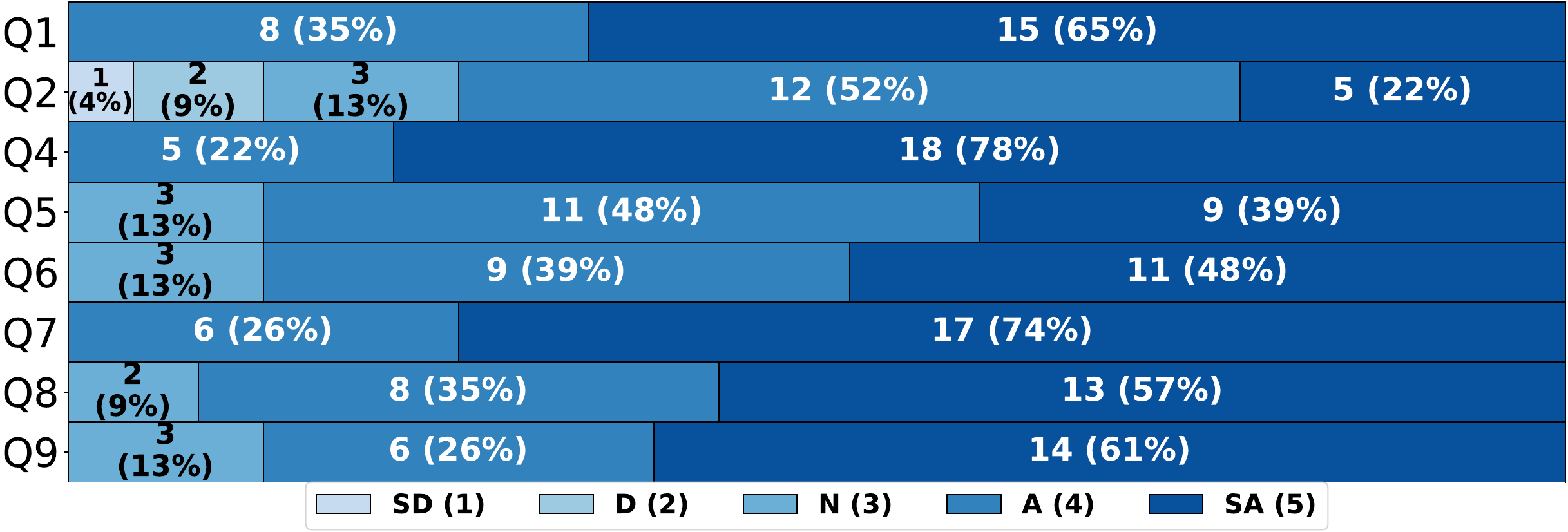}
    \caption{Likert-scale rating distribution for RQ3.}
    \label{fig:rq3-LIKERT}
\end{figure}

Figure~\ref{fig:rq3-LIKERT} reports the distribution of Likert-scale responses for all closed-ended survey items considered for RQ3.

Q1 received full positive agreement, with 8 \emph{A} and 15 \emph{SA}, indicating that practitioners perceived the report as useful for understanding the root cause of architectural drift. 
Q2 shows the most dispersed distribution, with 1 \emph{SD}, 2 \emph{D}, 3 \emph{N}, 12 \emph{A}, and 5 \emph{SA}. 
This suggests that, although the report supports drift interpretation, relating its information to the executed test cases may still require additional effort.
This limitation is consistent with the current design of the reporting component, which operates on the serialized delta and development-story summaries, but does not directly exploit full test specifications or explicit traceability links.
Q4 and Q7 received full positive agreement, with 5 \emph{A}/18 \emph{SA} and 6 \emph{A}/17 \emph{SA}, respectively, showing that participants perceived the methodology as effective in accelerating discrepancy identification and intuitive in supporting the interpretation of runtime message-exchange deltas.
Q5 obtained 3 \emph{N}, 11 \emph{A}, and 9 \emph{SA}, suggesting that the generated artifacts are considered useful for safety-oriented documentation, although additional traceability information is needed to better support ISO~26262 audit needs. 
Similarly, Q6 obtained 3 \emph{N}, 9 \emph{A}, and 11 \emph{SA}; Q8 obtained 2 \emph{N}, 8 \emph{A}, and 13 \emph{SA}; and Q9 obtained 3 \emph{N}, 6 \emph{A}, and 14 \emph{SA}. 
These results indicate that the language of the report, the system interface, and the potential regular use of the methodology in a CI/CD pipeline were generally evaluated positively.

The open-ended item Q3 asked participants whether they noticed discrepancies or hallucinations in the generated report with respect to the technical delta. 
Among the respondents, 20 reported none, while three mentioned minor wording oversimplifications or occasional ambiguity in root-cause descriptions. 
No respondent reported factual fabrications or omissions of delta content. 
This feedback suggests that the constrained reporting strategy limits hallucinations at the report level, while the remaining issues mainly concern the precision and contextual articulation of the explanation. The open-ended items Q10 and Q11 provided further insight into perceived value and desired improvements. 
Responses to Q10 mainly emphasized the automatic classification of drift root causes and the generation of a high-level, audit-oriented narrative that reduces the need to manually correlate multiple sources. 
Several respondents also valued the possibility of relating architectural changes to the development artifacts that triggered them. Responses to Q11 identified explicit traceability mapping as the main improvement area, especially from each reported drift to ISO~26262 work products, safety goals, and verification artifacts. 
Secondary requests included links to stories and commits, monitoring of architectural conformance trends over time, versioning metadata, and export capabilities including ASIL classification and mitigation status. 
A minority of participants indicated that no additional feature was needed, consistently with the overall positive Likert-scale ratings.

\begin{tcolorbox}[colback=gray!10,boxrule=0.5pt,title=RQ3 Answer,boxsep=1pt,left=1pt,right=1pt,top=1pt,bottom=1pt]
Practitioners perceive the generated artifacts as useful for interpreting architectural drift and supporting safety-oriented engineering workflows. 
The report is considered helpful for understanding root causes, accelerating discrepancy identification, and providing an audit-oriented narrative. However, strengthening the traceability between the report, the executed feature-linked test cases, development artifacts, and ISO~26262 work products remains the main area for improvement.
\end{tcolorbox}
\color{black}

\section{Threats to Validity}
\label{sec:threats}
This study proposes an automated methodology for detecting architectural drift by comparing design-time specifications with execution-level evidence. As with any empirical study, the proposed approach is subject to potential threats to validity. The main threats and mitigation strategies are discussed below.

\noindent \textbf{\textit{Construct validity.}}

The proposed methodology formalizes architectural drift as discrepancies between design-time and runtime interactions represented through sequence diagrams. Accordingly, the detected deltas capture only interaction-level deviations, classified as \textsc{confirmed}, \textsc{missing}, \textsc{additional}, or \textsc{inverted}. Other forms of drift, such as timing, data-flow, hardware-related, or asynchronous deviations, are outside the current scope. Since runtime behavior is reconstructed from function-call traces, interactions not materialized as explicit calls may remain unobserved. To mitigate this threat, design-time and runtime views are compared at the same abstraction level, and the resulting deltas are treated as candidate drift requiring expert review rather than complete architectural compliance evidence.

\noindent \textbf{\textit{Internal validity.}}
The correctness of the generated runtime view depends on the accuracy of binary metadata, such as DWARF information, used to resolve HIT traces into architectural entities. Imprecise metadata or symbol resolution may affect the inferred interactions, while predefined test procedures may not cover all execution paths. To mitigate these threats, the study relies on production-grade builds with full debug information and safety-oriented test procedures. The LLM-based step is confined to reporting already-computed deltas and does not affect trace processing, normalization, or delta computation.

\noindent \textbf{\textit{External validity.}} \label{external-validity}
The evaluation was conducted within a single industrial organization and on proprietary safety-critical firmware modules. Although the selected test cases reflect realistic industrial scenarios, the results may not generalize to other organizations, domains, architectures, or systems with highly concurrent, interrupt-driven, or non-deterministic execution models. The practitioner survey may also be affected by social desirability, common method bias, and evaluation apprehension, as participants belonged to the same industrial context in which the methodology was assessed. To mitigate these threats, participation was voluntary and anonymous, and open-ended questions were included to capture critical feedback beyond Likert-scale ratings. Further studies on additional systems, organizations, and execution models are needed to strengthen the generalizability of the findings.

\section{Lessons Learned from Industrial Adoption}
\label{sec:lessons}

This section discusses lessons learned from the industrial application of the proposed approach, drawing on the experimental evaluation and practitioner feedback.

\textbf{LL1: Algorithmic computation and LLM assistance address different failure modes.}

The proposed methodology separates the \emph{Differ} engine (Phase~2) from the LLM-assisted reporting layer (Phase~3) to address their complementary weaknesses. The Differ ensures reproducible, auditable classification of interactions, while the LLM layer reduces the cognitive load of interpreting the resulting delta. Reversing these roles would introduce variability and fabrication risks into safety-critical traceability chains. This suggests pipelines are a robust pattern for safety-oriented engineering methodology.

\textbf{LL2: Complementary representations support distinct engineering roles.}

The methodology produces two aligned outputs: a \emph{Serialized Delta} in PlantUML intended for automated processing, tooling integration, and traceability; and a natural-language report (\emph{Delta Report}) generated to support human review activities. Neither representation is sufficient in isolation.
The Serialized Delta provides precision and processability, but offers limited contextual support for architectural interpretation. Conversely, the Delta Report improves readability and reduces the interpretive effort during architectural reviews, but is inherently descriptive and less effective at explaining root causes, particularly in the absence of explicit test specifications and upstream traceability links, a limitation also reflected in survey feedback.
Practitioners consistently emphasized that LLM-generated reports are valuable only as supporting material. They do not replace design artifacts, verification evidence, or safety arguments required by ISO~26262, and must always be validated through expert review. Human-readable and machine-readable outputs therefore serve complementary roles at different stages of the engineering and review workflow.

\section{Related Work}
\label{sec:related}

Research related to our work connects architecture recovery and documentation for embedded and safety-critical automotive software with trace-based runtime analysis, runtime verification, and the use of execution traces as behavioral evidence in software engineering.

Software architecture recovery has been widely studied to reconstruct architectural knowledge from implementation artifacts. Source-code- and static-analysis-based approaches can recover architectural views of automotive embedded software, support communication-level consistency checks, and generate UML documentation such as package, component, component \& connector, and state-machine diagrams~\cite{zhang2014,amalfitano2024}. These works are valuable when documentation is incomplete or outdated, but remain mainly static: they infer structures, dependencies, and design-level relations from code, without capturing which interactions are exercised in a specific runtime scenario. This limitation reflects a broader challenge in embedded-system reverse engineering, where hardware/software interplay, timing, safety criticality, and the integration of static and dynamic information must be considered~\cite{kienle2012software}. Architecture recovery has also been highlighted as a means to understand legacy ECU software and support verification and design-space exploration~\cite{zhang2017automated}.

Accurate architectural information is central to safety-critical automotive software. Prior work highlights the difficulty of producing complete ISO~26262-aligned architecture documentation~\cite{amalfitano2023}, discusses compliance-oriented modeling and planning practices such as testability, traceability, freedom from interference, and partitioning~\cite{Gross2020}, and surveys automotive reference architectures, including AUTOSAR, view-based modeling, and abstractions for managing complexity and safety requirements~\cite{bauer2022}. These studies frame architecture as essential for safety reasoning and compliance, but focus on how it should be designed, organized, or documented, rather than on how executed firmware behavior can be recovered and projected onto architectural views. Runtime evidence is central to trace-based analysis and runtime verification of embedded and real-time systems. Prior work presents tracing as a practical mechanism for understanding runtime behavior in resource-constrained platforms and diagnosing performance or interaction issues~\cite{chamski2010}, and proposes hardware-based runtime verification to reconstruct control-flow traces and evaluate temporal properties online through FPGA-based processing~\cite{convent2018}. These works provide accurate execution evidence with limited interference, but target runtime checking, timing analysis, or low-level observation rather than architectural runtime views or comparisons with statically recovered architectures. A related conceptual foundation is the broader \emph{$Models@run.time$} area, surveyed by Bencomo \emph{et al.}~\cite{bencomo2019models}, where running systems remain causally connected to architectural abstractions to support runtime reasoning, adaptation, and verification. While close to our goal of producing an executed architectural view, full \emph{$Models@run.time$} frameworks require continuous synchronization, which may introduce unsustainable overhead in resource-constrained firmware. Our approach instead adopts lightweight offline trace-based reconstruction and delta analysis, elevating runtime evidence to the architectural level without continuous synchronization. Recent studies further show that execution traces can enrich software understanding tasks. Haque \emph{et al.}~\cite{haque2025traces} and Wu \emph{et al.}~\cite{wu2026runtime} investigate execution traces with LLMs, showing that runtime traces can provide behavioral evidence beyond static code, although their usefulness depends on trace representation and management. While these works are not concerned with architecture recovery, they reinforce the idea that traces capture behavioral information that static artifacts may miss. In our work, however, traces are not used to repair code or guide an LLM's defect reasoning, but as direct evidence to reconstruct an executed architectural view of firmware behavior.

LLMs have also been explored for architectural understanding and documentation. Hatahet \emph{et al.}~\cite{hatahet2025generating} propose a semi-automated pipeline that combines static reverse engineering and LLM-based abstraction to generate architectural descriptions from source code, including structural and behavioral views. In contrast, our approach does not delegate architectural inference to the LLM: relations and deltas are computed deterministically from design artifacts and runtime traces, while the LLM only post-processes pre-computed deltas to improve human interpretability.

Overall, prior work separates static architecture recovery and documentation, which provide structural views for embedded and automotive software~\cite{zhang2014,amalfitano2024,amalfitano2023,bauer2022}, from trace-based analysis and runtime verification, which provide fine-grained evidence of executed behavior~\cite{chamski2010,convent2018,kienle2012software}. Their integration into runtime-informed \emph{architectural} documentation for firmware systems remains less explored. Our work addresses this gap by transforming HIT runtime traces into architectural diagrams of observed interactions and relating this executed view to the statically recovered architecture through delta analysis. Rather than replacing static recovery or runtime tracing, the proposed workflow connects them to support architecture comprehension, documentation alignment, and safety-oriented reasoning in embedded firmware.

\section{Conclusions and Future Work}
\label{sec:conclusion}

This paper presented a runtime-informed methodology for detecting architectural drift in safety-critical firmware by comparing design-time behavioral specifications with runtime evidence. The approach produces an explicit interaction-level delta between intended and observed behavior, while a constrained LLM-based step is used only to generate a human-readable report supporting expert review. In this way, the methodology preserves deterministic drift detection while improving the interpretability of the resulting discrepancies. The industrial evaluation suggests that the proposed methodology can support the identification and interpretation of design-runtime discrepancies in ISO~26262-compliant firmware. The validation results show strong alignment between the generated deltas and expert-curated references, especially for confirmed, missing, and additional interactions, while order-related discrepancies remain more challenging. Practitioner feedback further indicates that the generated artifacts are perceived as useful for reducing manual inspection effort and supporting safety-oriented documentation activities. At the same time, the LLM-generated reports should be considered review aids rather than primary safety evidence. These findings should be interpreted within the scope of the evaluated industrial context and the interaction-level notion of drift adopted in this work. Future work will extend the evaluation to additional systems and organizations, improve the handling of ordering-related discrepancies, investigate the extension of the differ to UML combined fragments such as \texttt{alt}, \texttt{opt}, and \texttt{loop}, and strengthen traceability between detected drift, feature-linked test cases, development artifacts, and ISO~26262 work products.

\section*{Acknowledgment}

This work is an independent academic contribution and does not represent the views of Micron Technology, Inc. or its affiliates. The proposed methodology relies exclusively on publicly available tools, open-source libraries, and established software engineering techniques. \emph{Enterprise Architect} is cited only as a representative architectural modeling environment, without implying endorsement or dependency on a specific tool~\cite{Manteuffel2016_EA, Zimmermann_EA}. No proprietary data, confidential information, or industrially developed algorithms, tools, or implementations are used or disclosed.

\clearpage
\balance
\bibliographystyle{IEEEtran}
\bibliography{biblo}

\end{document}